\documentclass[twocolumn,aps]{revtex4}
\usepackage{graphicx,amsmath,epsfig} 
\usepackage{indentfirst,here}

\begin{document} 
   
\title{Friction vs Texture at the Approach of a Granular Avalanche }

\author{Lydie Staron$^1$ and Farhang Radja\"\i$^2$}

\affiliation{ $^1$ DAMTP, University of Cambridge, Wilberforce Road, Cambridge CB3 0WA, UK. \\ 
$^2$ {LMGC, CNRS-Universit\'e Montpellier II, Place~Eug\`ene~Bataillon,~F-34095~Montpellier~Cx,~Fr.}
}

\begin{abstract}  
We perform a novel analysis of the granular texture of a granular bed close to stability limit. Our analysis is based on a unique criterion of friction mobilisation in a simulated two-dimensional packing. In this way, we recover the bimodal character of granular texture, and the coexistence of weak and strong phases in the sense of distinct contacts populations. Moreover, we show the existence of a well-defined  subset of contacts within the weak contact network. These contacts are characterized by their important friction, and form a highly coherent population in terms of fabric. They play an antagonistic role with respect to force chains. We are thus able to discriminate between incoherent contacts and coherent contacts in the weak phase, and to specify the role that the latter plays in the destabilisation process.
\end{abstract}
\maketitle 

\begin{figure}
\centerline{\includegraphics[angle=-90, width=0.90\linewidth]{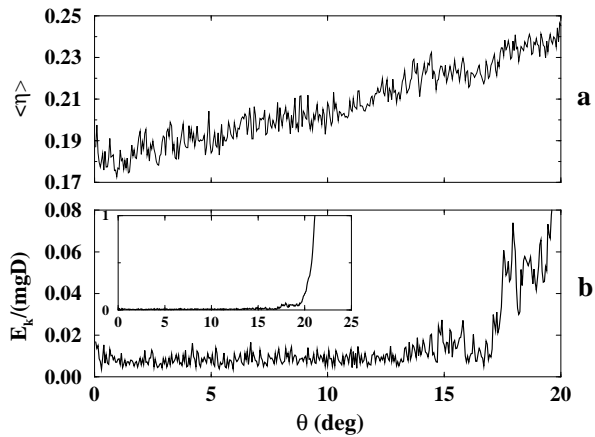}}
\caption{a) Mean mobilisation of friction forces $\langle \eta \rangle_{N_c}$ averaged over all the contacts and b) mean kinetic energy $E_k$ normalised by the typical potential energy of a grain of mass $m$ for a height $D$ as a function of the slope angle $\theta$.}
\label{Fig1}
\end{figure}

\begin{figure}
\centerline{\includegraphics[angle=-90, width=0.90\linewidth]{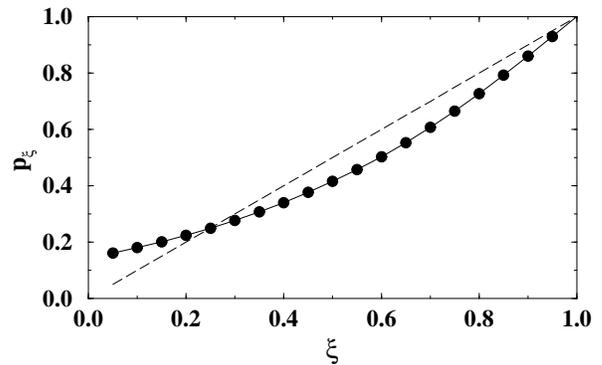}}
\caption{Proportion of contacts $p_\xi$ belonging to the subset of contacts $C_\xi$. The dashed line represents the case of a uniform distribution of $\eta$.}
\label{Fig2}
\end{figure}

\begin{figure}
\centerline{\includegraphics[angle=-90, width=0.9\linewidth]{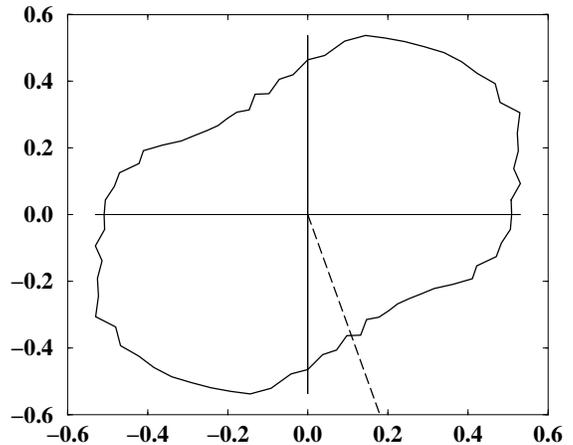}}
\caption{Polar representation of the mean friction mobilisation $\eta_\varphi$ as a function of the orientation $\varphi$ of the contacts.}
\label{etapolar}
\end{figure}

Disordered systems exhibit critical behaviours which often result in the emergence of rare ``catastrophic'' events for which predictive modelling is problematic. Earthquakes, and more generally, rupture of composite materials, are well known example of this criticality~\cite{andersen97}. Dry cohesionless granular matter,  characterized by a generic disorder induced by the rigidity of the grains and the highly dissipative contacts between them, obeys the same type of behaviour. Its response to loading involves a succession of micro-plastic events which eventually lead to a catastrophic failure~\cite{volfson04,darve99}, for instance, a surface avalanche.\\  
In spite of their intrinsic disorder, granular materials obey a well-defined organisation commonly referred to as {\em granular texture}~\cite{radjai98b}. In particular, only a fraction of the contacts contributes to the quasi-totality of the shear stress. This fraction, known as the {\em strong contact network} and forming {\em force chains}, has been extensively studied on the ground that it is responsible for the mechanical strength of the whole granular structure. The organisation of frictional contacts, {\it i.e.} the dissipative part of the contact network, has attracted less attention. Yet they are responsible for the local dynamics observed before failure and are essential for the identification of precursors of destabilisation~\cite{staron02,kabla}.\\
In this Letter, we propose an analysis of granular structure based solely on the criterion of friction mobilisation. Doing so we are able to recover the bi-phasic nature of granular texture. More importantly, we show for the first time the   existence of a well defined population of contacts, highly frictional and coherent, and coexisting in the weak phase with a large population of incoherent contacts. These findings enable us to revisit the characteristics of the granular texture, and the respective roles of the weak and the strong contacts in the destabilisation process.\\
  This analysis was performed by means of two-dimensional numerical simulation. We have used a contact dynamics algorithm~\cite{jean_moreau92} assuming perfectly rigid grains interacting at contacts through a hard core repulsion and a Coulomb friction law. Beyond the fact that contact dynamics treats them as strictly non-smooth, these contact laws do not differ from those more commonly used in discrete simulations~\cite{allen87}.\\
We consider a granular bed built by random rain of $8000$ circular grains of diameters varying in a ratio $1.5$ to induce geometrical disorder. The solid fraction of the packing is $0.82$. The height of the bed is $40D$ and its width is $200D$, where $D$ is the grain mean diameter. To prevent rigid boundary effects, the bed is periodic in the horizontal direction, namely no vertical walls are introduced to confine the packing. A constant rotation rate is applied (corresponding to a rotation of $10^{-3}$ deg per time step), gradually bringing the slope $\theta$ of the bed from initially zero to the limit value $\theta_c$ for which a surface avalanche is triggered. \\
The Coulomb friction law imposes an upper limit on the tangential force  ${f_T}$ at contact. This upper limit (the Coulomb threshold) is given by $\mu f_N$, where ${f_N}$ is the normal force at contact and $\mu$ is the coefficient of friction. A necessary condition for slip motion to occur between two grains in contact is ${f_T}=\mu f_N$. In the following, the value of $\mu$ is set to $0.5$ and its influence, though not investigated here, will be discussed later. A coefficient of restitution allows for the modelling of energy exchanges during binary collisions; since we are interested in micro-plasticity only, we set its value to zero and consider purely dissipative contact interactions. \\
The mobilisation of friction at each contact is simply measured by the ratio of the tangential to the normal force times the coefficient of friction: 
\begin{equation}
\eta = \frac{1}{\mu}\frac{f_{T}}{f_{N}},
\label{eta}
\end{equation}
with $\eta \in [0,1]$ following the friction law. The case $\eta = 1$ coincides with the Coulomb threshold and a probable slip motion at the contact. We compute the mean value $\langle \eta \rangle$ over the total number of contacts $N_c$ in the packing as a function of the slope angle $\theta$ (Fig.~\ref{Fig1}a). The mean normalised kinetic energy of the grains as a function of $\theta$ is displayed in Fig.~\ref{Fig1}b. The onset of the avalanche occurs for $\theta=\theta_c\simeq 20^\circ$. While the mean level of friction increases regularly, we observe some dynamical activity intensifying few degrees before $\theta_c$. This interval corresponds to a metastable state during which local dynamical rearrangements occur more frequently due to an enhanced mobilisation of friction~\cite{daerr99,deboeuf03}. In the following, we thus focus on this interval of slope angles $\Delta_\theta = [\theta_c - 3^\circ, \theta_c]$, and the quantities discussed hereafter are averaged over it.\\
We consider now different populations of contacts following their friction level $\eta$. We introduce a parameter $\xi$ to which the friction level $\eta^\alpha$ is compared for each contact $\alpha$. A contact $\alpha$ belongs to $C_\xi$ if $\eta^\alpha \ge (1-\xi)$: 
\begin{equation}
C_\xi = \{\alpha \in N_c \:/\: \eta^\alpha \ge (1-\xi)\}.
\end{equation}
 The parameter $\xi$ varies in the interval $[0, 1]$. The number of contacts belonging to $C_\xi$ is $N_\xi$. The proportion  $p_\xi = N_\xi/N_c$ of contacts belonging to $C_\xi$ is plotted against $\xi$ in Fig.~\ref{Fig2}. 
For $\xi$ close to $0$ ({\it i.e.} for a high mobilisation of friction), $p_\xi$  reaches $18\%$, exceeding by far the case of a uniform distribution. This value bespeaks the existence of a local peak in the distribution of $\eta$. Contacts such that $\xi \rightarrow 0$ thus form a well defined subset. In the same way, the increase of $p_\xi$ when $\xi \rightarrow 1$ is more rapid than a uniform distribution would allow, showing again the non-uniformity of the distribution of friction at contacts. We have checked that these features were robust with respect to $\mu$ and the boundary conditions in piles close to stability limit.\\
A picture of the mobilisation of friction depending on contact orientation is obtained when plotting the value of $\eta$ averaged over angular sectors $\varphi + \Delta \varphi$ for contacts directions:
\begin{equation}
\eta_\varphi = \frac{1}{N_c^\varphi} \sum_{\alpha=1}^{N_c^\varphi} \eta^\alpha,
\end{equation}
where $N_c^\varphi$ is the number of contacts for which the orientation of the unit vector $\boldsymbol{n}$  normal to the contact plan belongs to $[\varphi, \varphi+\Delta \varphi]$. The function $\eta_\varphi (\varphi)$ is plotted in Fig.~\ref{etapolar} in  polar coordinates. The slope of the granular bed coincides with the horizontal direction, and the direction of gravity is represented by a dashed line (at~$-70^\circ$). We observe a higher mobilisation of friction for contacts directed at $\simeq 25^\circ \:[\pi]$, {\it i.e.} nearly  perpendicular to the gravity. Note that this angular map of friction mobilisation does not need to compare with the angular map of contact orientations.\\ 
\begin{figure}
\centerline{\includegraphics[angle=-90, width=0.99\linewidth]{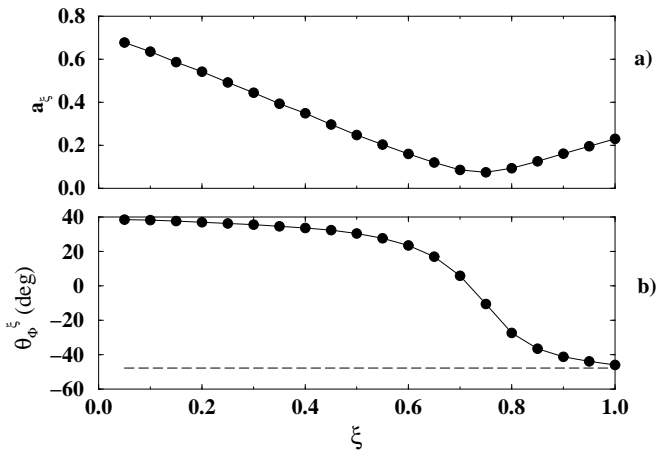}}
\caption{ a) Anisotropy $a_\xi$ of the contact network in the subset $C_\xi$
and b) major fabric direction ${\theta}^{\xi}_{\Phi}$ (counted positively anticlockwise) as a function of $\xi$ (see text). The dashed line shows the major stress direction over all the contacts.}
\label{mutext}
\end{figure}
To compare the anisotropy of friction mobilisation with the anisotropy of the geometrical contact network, we compute {\em partial fabric tensors} $\bar{\Phi}^\xi$ defined over the contact subsets $C_\xi$~\cite{satake82}:
\begin{equation}
\Phi^\xi_{ij} = \frac{1}{N_\xi} \sum_{\alpha = 1}^{N_\xi} n_i^\alpha n_j^\alpha,
\end{equation}
where $\boldsymbol{n}^\alpha$ is the unit vector of contact $\alpha$, and $i$ and $j$ refer to coordinates $x$ and $y$ (with $x$ parallel to the free surface, and $y$ perpendicular to the free surface and pointing downward). The major principal direction $\theta_\Phi^\xi$ of the tensor $\bar{\Phi}^\xi$ corresponds to the mean direction of the contacts in $C_\xi$ regarding the slope. The difference of the eigenvalues gives the anisotropy $a_\xi$ of the contact sub-network $C_\xi$. These two quantities are displayed in Fig.~\ref{mutext} against $\xi$. We see that the largest anisotropy occurs for $\xi \simeq 0$, {\it i.e.} for high friction mobilisation. It then declines regularly and passes by its lowest value (close to zero) for $\xi \simeq 0.7$ before increasing again up to $\xi \simeq 1$. This transition coincides with a rotation of $\pi/2$ of the fabric tensor.\\
 We can thus discriminate between three groups of contacts on the unique criterion of friction mobilisation:
\begin{enumerate}
\item Contacts  with a very high mobilisation of friction ($\xi \rightarrow 0$), {\it i.e.} contacts where slip motion is likely to happen. Their orientations obey a maximum anisotropy ($0.75$), three times superior to the anisotropy of the whole packing, and define a mean direction at $\simeq 40^\circ$, orthogonal to the major stress diection ($\simeq -50^\circ$). These contacts are thus forming a highly {\em coherent} subset in terms of fabric.
\item Contact with intermediate level of friction mobilisation ($0.1< \xi <0.7$): they very weakly contribute to the geometrical anisotropy. Accordingly their mean orientation allows only for a small deviation from the direction dictated by the first group of contacts. Their small contribution to the fabric suggests that they form an {\em incoherent} ensemble.  
\item Contact with a very low mobilisation of friction ($\xi \rightarrow 1$): they strongly contribute to the geometrical anisotropy. Accordingly, their direction allows for a dramatic rotation of the fabric tensor, forcing the latter to follow the major stress direction.  Their strong contribution to the fabric show that they form a {\em coherent} ensemble. They represent $40\%$ of the total number of contacts (see Fig.~\ref{Fig2}).
\end{enumerate}

\begin{figure}
\centerline{\includegraphics[angle=-90, width=0.90\linewidth]{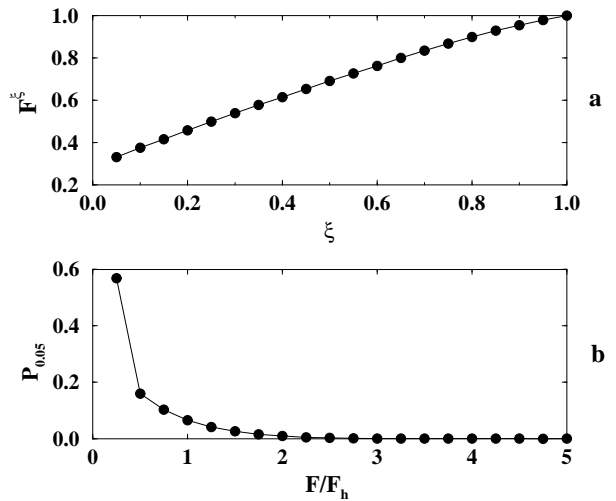}}
\caption{a) Mean force $F^\xi$ transmitted by contacts of $C_{\xi}$ as a function of $\xi$ and b) Probability density function $P_{0.05}$ of $F/F_h$ within the subset of contacts $C_{0.05}$.}
\label{wsvsxi}
\end{figure}
Discriminating between contacts on the ground of contact friction discloses features strongly reminiscent of the bimodal character of granular texture~\cite{radjai98b}.
 The granular texture is characterized when discriminating between contacts on the ground of the intensity of the force they transmit. In the case of our granular bed, the effect of gravity has to be filtered out to distinguish between weak and strong contacts. This is done by normalizing all contact forces $F$ by the mean contact force $F_h$ at the depth $h$ corresponding to the position of the contact. The mean value $F^\xi$ of the normalised force transmitted by the contacts of $C_\xi$ is plotted in Fig.~\ref{wsvsxi}a as a function of $\xi$: higher mobilisation of friction force (small $\xi$) coincides with weaker forces and {\it vice versa}. This concordance between force transmission and friction mobilisation implies that the transitions observed in Fig.~\ref{mutext} partly reflect the transition from the weak to the strong contact network.\\
Focusing on contacts for which the mobilisation of friction is the highest, we consider the subset $C_{0.05}$, {\it i.e.} the contacts satisfying $\eta \ge 95\%$. For these contacts, we plot in Fig.~\ref{wsvsxi}b the distribution of the intensity $P_{0.05}(F/F_h)$ of the forces they transmit. We observe that most contacts of $C_{0.05}$ transmit very weak forces, such that $F/\langle F\rangle_h \le 0.25$. In any case, $90\%$ of the contacts of $C_{0.05}$ belong to the weak contact network, namely transmit forces such that $F/F_h<1$.\\
\begin{figure}
\centerline{\includegraphics[bbllx=270,bblly=100,bburx=400,bbury=200,clip=, width=0.9\linewidth]{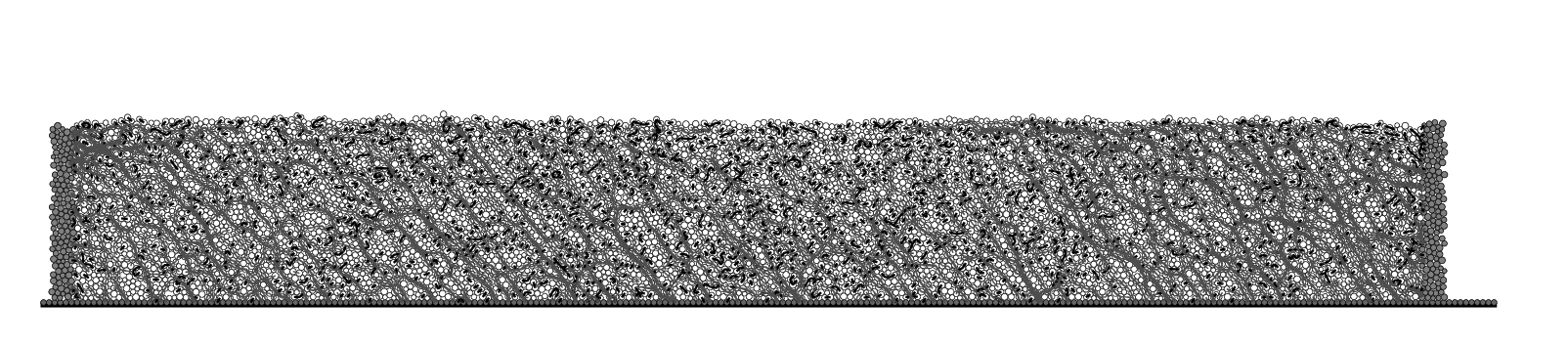}}
\caption{Sample of the granular bed in the interval $[\theta_c -3^\circ, \theta_c]$. In gray are represented the force chains (contacts such that $F/F_h >1$) and in black contacts such that $\eta \ge 0.95\mu$.}
\label{texture}
\end{figure}
The interplay between force transmission and friction mobilisation is illustrated in Fig.~\ref{texture}, where one snapshot of the state of the granular pile at an instant of the interval $\Delta_\theta = [\theta_c - 3^\circ, \theta_c]$ is shown. In gray are represented the force chains defined by the strong contact network, with line width varying proportionally to the force intensity. Black segments of fixed width represent the contacts belonging to $C_{0.05}$. The following features are apparent: contacts of $C_{0.05}$ are belonging to the weak phase (namely none of them is part of a force chain), they show a strong geometrical anisotropy (for we distinguish indeed a leading orientation),  and their direction is orthogonal to the main stress orientation ({\it i.e.} contacts of $C_{0.05}$ are in their great majority transverse to force chains). \\

{\em On the unique criterion of friction mobilisation} (as defined by equation~(\ref{eta})) in a packing close to avalanching, we are able to recover the bimodal character of granular packing and the existence of two distinct phases, in agreement with the analysis in terms of a strong and a weak contact network. Moreover, we show {\em the existence of a well-defined population of contacts within the weak contact network}. These contacts are defined by their very high mobilisation of friction. They form a highly coherent population, showing a strong anisotropy and a well defined mean direction. In this sense, they strongly differ from other weak contacts. Their orientation, orthogonal to force chains, and their susceptibility to slip due to their level of friction, impart to them an antagonist role with respect to force chains. Frictional contacts and strong contacts form the extremal state of our analysis, and they play extremal roles in terms of stability. By contrast, all other contacts in the weak network are highly incoherent.\\
This analysis relies on the relative value of the normal and tangential force at contacts, and is independent of any mean quantity or any cut-off value reflecting the mean state of the packing. As such, it purely reflects the intrinsic organisation of the packing. However, the level of friction mobilisation is necessarily measured with respect to the coefficient of friction $\mu$. In particular, the proportion of highly frictional contacts is expected to decrease when the value of $\mu$ increases. An investigation of the effect of this parameter on the results reported here should be undertaken. However, in the same way that the characteristics of force transmission are robust with respect to the material properties, we expect the organisation of friction to be a robust feature of granular texture. \\ 
Back to the description in terms of strong and weak contact networks, the analysis of granular texture in terms of friction allows to revisit the role of these two entities. The contribution of the strong contact network to the stress state designates it as solid skeleton of the granular structure, and responsible for its stability. By contrast, the weak contact network appears as an ``interstitial'' phase essentially screened from external loading, and proping the force chains. In this work, we show that the weak contact network is actually hosting a subset of contacts whose role in destabilisation is fundamental. Although forming a less coherent phase when analysed on the ground of force transmission ({\it i.e.} representing a small anisotropy), the weak contact network contains an intrinsic coherence when analysed on the ground of friction. This makes the weak contact network an active agent of the destabilisation process. In other words, the macroscopic behaviour of the granular packing is partly dependent on a phenomenology which escapes the mean description in terms of stress state.\\
This work was supported by the Marie Curie European Fellowship FP6 program. 

\end{document}